# Generation of narrowband quantum emitters in hBN with optically addressable spins

Benjamin Whitefield*,[1,2], Helen Zhi Jie Zeng*,[1], James Liddle-Wesolowski[1,2], Islay O. Robertson[3], Viktor Ivády[4,5], Kenji Watanabe[6],Takashi Taniguchi[7], Milos Toth[1,2], Jean-Philippe Tetienne[3], Igor Aharonovich[1,2] and Mehran Kianinia[†,1,2]

[1] School of Mathematical and Physical Sciences, University of Technology Sydney, Ultimo, New South Wales 2007, Australia
[2] ARC Centre of Excellence for Transformative Meta-Optical Systems, University of Technology Sydney, Ultimo, New South Wales 2007, Australia
[3] School of Science, RMIT University, Melbourne, VIC 3001, Australia
[4] HUN-REN Wigner Research Centre for Physics, P.O. Box 49, H-1525 Budapest, Hungary.
[5] MTA–ELTE Lendület "Momentum" NewQubit Research Group, Pázmány Péter, Sétány 1/A, Budapest, 1117, Hungary
[6] Research Center for Electronic and Optical Materials, National Institute for Materials Science, 1-1 Namiki, Tsukuba 305-0044, Japan
[7] Research Center for Materials Nanoarchitectonics, National Institute for Materials Science, 1-1 Namiki, Tsukuba 305-0044, Japan305-0044, Japan*These authors contributed equally

*These authors contributed equally

† To whom correspondence should be addressed: mehran.kianinia@uts.edu.au; igor.aharonovich@uts.edu.au

## Abstract

Electron spins coupled with optical transitions in solids stand out as a promising platform for developing spin-based quantum technologies. Recently, hexagonal boron nitride (hBN) - a layered Van der Waals (vdW) crystal, has emerged as a promising host for optically addressable spin systems. However, to date, on-demand generation of isolated single photon emitters with pre-determined spin transitions has remained elusive. Here, we report on a single step, thermal processing of hBN flakes that produces high density, narrowband, quantum emitters with optically active spin transitions. Remarkably, over 25% of the emitters exhibit a clear signature of an optical spin readout at room temperature, surpassing all previously reported results by an order of magnitude. The generated spin defect complexes exhibit both $S = 1$ and $S = 1/2$ transitions, which are explained by charge transfer from strongly to weakly coupled spin pairs. Our work advances the understanding of spin complexes in hBN and paves the way for single spin - photon interfaces in layered vdW materials with applications in quantum sensing and information processing.

## Introduction

Optically active spin defects in solids are highly sought after for quantum technologies, including quantum communication, quantum networks, and quantum sensing[1-3]. In solids, optical detection of spin states has been realised for a variety of point defects such as nitrogen-vacancy (NV) and silicon-vacancy (SiV) in diamond, divacancy in silicon carbide (SiC)[4,5], defects in hexagonal boron nitride (hBN) and gallium nitride (GaN)[6,7]. Among these, spin defects in hBN have garnered significant attention with the zero phonon line (ZPL) emissions spanning both visible and near-infrared regions[8]. Their unusual spin dynamics have driven fundamental research, presenting promising opportunities for advancing spin-based technologies using two-dimensional materials.

The first room-temperature spin defect in hBN was identified as the negatively charged boron vacancy ($V_B^-$) with broad spin triplet transitions that are hyperfine coupled to three nearby nitrogen atoms[9]. Boron vacancies were localised within few-layer hBN flakes[10], nanoflakes and nanotubes and used for a range of sensing applications[11,12]. However, they have only been observed in ensemble measurements to date and isolating a single $V_B^-$ defect remains a challenge due to its low quantum efficiency[13].

Alternatively, a group of single photon emitters in hBN have been observed by various groups with a complex spin system at room temperature[6,7]. The photophysics around their spin multiplicity remains under debate, with some groups observing S = 1 systems[14] whilst most report S = 1/2 behaviour. Recently the optically detected magnetic resonance (ODMR) mechanism for the S = 1/2 system has been explained in the form of charge transfer within between two nearby defects forming a weakly coupled spin pair[15]. All single defect - single spin systems were, however, detected in sub-optimal hBN samples such as particles and polycrystalline films, and never engineered on demand in exfoliated flakes[6,16-19]. Exfoliated flakes are key for realisation of integrated photonic devices, and the ability to engineer spin defects in flakes is therefore of paramount importance.

Despite the significant progress in identifying single spin defects in hBN, exploration and utilization of these spin defects in quantum technologies remain largely untouched primarily due to challenges in deterministic fabrication of the defects and suboptimal spectral properties. In as-grown hBN films and treated hBN powders or flakes, the probability of identifying a single emitter with spin transitions is below 5%, with only 1 in 20 emitters exhibiting optically active spin transitions. Moreover, the spectral emissions recorded from these emitters in previous reports are broad, making them less appealing for quantum applications.

In this study, we controllably engineer spin defects in multilayer pristine hBN flakes and carbon-doped hBN (c-hBN) using an oxygen-annealing process. Our results demonstrate a robust protocol with nearly an order of magnitude improvement over previous attempts to generate optically active single-spin defects in hBN with narrow emission lines. Moreover, we show that the population of spin defects in c-hBN is five times higher compared to spin defects in pristine hBN. These findings are significant as they address the long-standing issue of achieving narrowband emitters with spin-active properties and demonstrate a viable method for engineering them in hBN.

To generate a high density of spin-active emitters in hBN, We begin by exfoliating thin (100-250 nm) hBN flakes from two types of hBN crystals. Pristine hBN (referred to as hBN in this work) and carbon-doped hBN (referred to as c-hBN in this work). The c-hBN crystal was obtained by annealing the hBN crystal at elevated temperatures in the presence of graphite, resulting in the diffusion of carbon into the hBN crystal[20]. c-hBN was chosen as recent studies of spin defects and single photon emitters in hBN allude to the presence of carbon related defects[21-24]. Both hBN and c-hBN were exfoliated onto Si/SiO$_2$ substrate and annealed under oxygen at 1000 °C for 4 hours as shown schematically in Fig. 1 (a) (See Methods and Fig. S1 for further details).

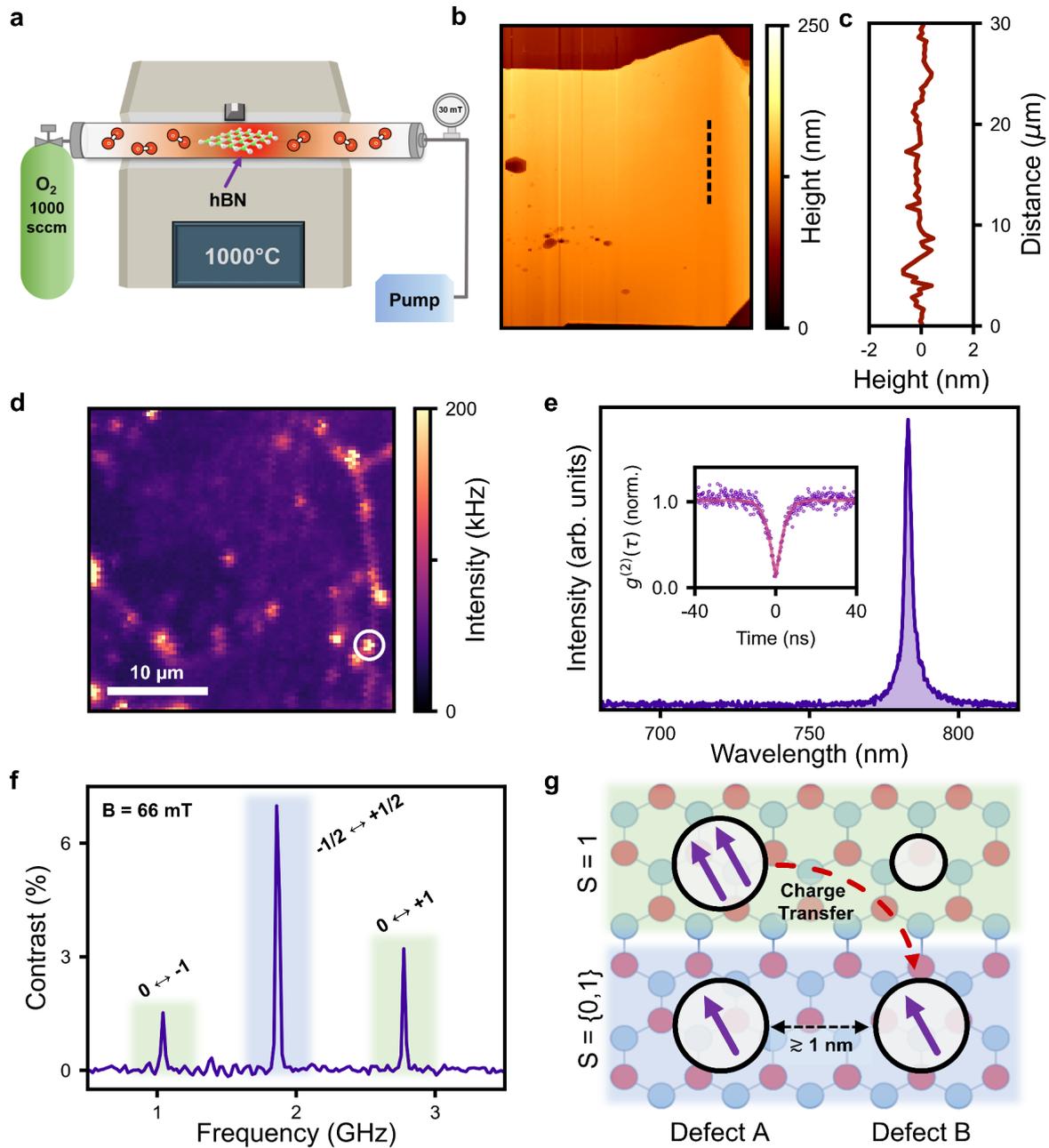

Figure 1. Generation of quantum emitters with optically addressable spins in hBN a) Oxygen annealing process at 1000 °C used to activate emitters in hBN flakes. b) Atomic Force Microscopy (AFM) scan of oxygen annealed hBN flake with a thickness of 140 nm c) height profile along the surface of the flake marked in (b) with a surface roughness of ±1 nm d) Confocal PL map of an oxygen annealed c-hBN showing spatially distinct emission hotspots. e) Characteristic room temperature photoluminescence spectrum of an oxygen annealed c-hBN emitter circled in c) with a zero phonon line at 781 nm. Inset displays the second-order correlation function $g^2(\tau)$ with a dip at $g^2(0)=0.1$, indicating single photon emission. f) An ODMR

*spectrum of a single emitter in c-hBN measured with an external magnetic field of 66 mT with two resonances at ~1.1 and 2.8 GHz from the $m_s$ = 0 ↔ ±1 transitions as well as a transition at ~1.9 GHz related to the S = 1/2 transition. g) Proposed spin complex responsible for the single photon emission as well as the spin - photon interface, consisting of two nearby defects A and B. Initially (top), Defect A can be excited to an S = 1 state. The transfer of an electron from Defect A to B changes both defects to an S = 1/2 state (bottom). The blue and green highlights indicate the resonance associated with each electron configuration.*

Thermal annealing under oxygen activates emitters across the visible spectral range with narrow linewidth at room temperature. To verify the quality of hBN flakes and emitters after the oxygen annealing process, Photoluminescnece (PL) and AFM characterization was performed. The surface of the hBN flake after oxygen annealing remains mostly intact, although some pinholes can be found on some areas of the flake as shown in Fig. 1 (b). These pinholes are created as the result of hBN interaction with oxygen at high temperatures, however, the remaining part of the flakes where emitters are found remains unetched as confirmed by AFM measurement. Further elemental analysis (Fig. S6) also did not show significant presence of oxygen or other elements into hBN after annealing.

Fig. 1 (c) is a measurement of roughness across the dashed line in Fig. 1 (b), indicating surface roughness of ~ 1 nm which is similar to the quality of the hBN crystals before annealing. After annealing, high density of emitters are generated within the hBN flake (refer to supporting information Fig. S2 -S4 for more PL characterisation of generated emitters). Fig. 1 (d) shows the confocal PL map of an oxygen annealed c-hBN flake with the location of a particular emitter in Fig. 1 (e) highlighted by a white circle. Note that many spectrally distinct hotspots corresponding to other quantum emitters are clearly visible in the confocal map. The emitters reported in this study were measured using a home-built confocal microscope (See Methods). The particular emitter in Fig. 1 (e) has a narrow zero phonon line (ZPL) at 781 nm and a FWHM of 2.76 nm  The quantum nature of the emission was examined by measuring the $2^{nd}$ order autocorrelation (Fig. 1 (e) inset). The dip of ~0.1 at zero delay time $g^{(2)}(τ) = 0$ indicates the emission is from a single quantum emitter.

The emitters created by this method have narrow zero phonon lines (ZPLs) and can be found across the visible spectral range with ZPLs spanning 580 to 850 nm (shown in Fig. S2). This broad spectral range is a significant advantage, as it provides a diverse set of emitters with varying emission wavelengths, offering flexibility for different quantum applications that may require specific wavelengths for integration with other systems. The ability to generate emitters across such a wide range also indicates the robustness of the oxygen-annealing process, which successfully induces a variety of optically active defects in the material. More importantly, most of the generated single emitters show narrow ZPL, typically less than 10 nm at room temperature (Fig. S5). The Narrow ZPLs are key indicators of high-fidelity single-photon emission, which is essential for applications requiring coherent single photons in addition to optically active spin transitions such as quantum networks and computing.

Fig. 1 (f) shows a representative optically detected magnetic resonance (ODMR) spectrum recorded from the quantum emitter in c-hBN in the presence of a 66 mT static magnetic field applied normal to the hBN flake (out-of-plane) (See Methods). Three resonances are visible which correspond to S = 1 (highlighted in green) with a zero-field splitting of D ≈ 850 MHz and

S = 1/2 (highlighted in blue) spin transitions. The $m_s = 0 \leftrightarrow \pm1$ transitions (S = 1) appear at ~1.1 and ~2.8 GHz respectively. As well as the S = 1/2 transition at ~1.9 GHz.

In contrast to previous reports[7,8,14-19] on similar carbon related systems in hBN which have only observed the S = 1/2 transition, the three resonances are observed within a single emitter (more data available in Fig. S10). We further note that we have never observed a quantum emitter with solely a S = 1 transition, and all defects that exhibited a S = 1 transition also showed a S = 1/2 transition. The reverse however, did occur and occasionally emitters with only S = 1/2 transitions were found (more details below and Fig. 2 (d)). Also, a similar spin complex with both S = 1 and S = 1/2 was found in hBN nanopowder and therefore they are not exclusive to exfoliated flakes, however, its occurrence in nanopowder appears to be rare or difficult to identify (see Fig. S9).

The spin dependent optical transitions for S = 1 systems are well understood and explained by photodynamic modelling with an imbalance in the transition rates between a metastable state and the $m_s = 0$ and $m_s = \pm1$ spin sublevels of the ground and excited states[25-27]. However, such models have not been successfully reconciled with the observed S = 1/2 transitions in a satisfying manner. Recently a charge-transfer model where a weekly coupled spin pair hosted by two independent defects (separated by a distance ≥1 nm distance) has been proposed for the optical detection of the S = 1/2 transition[15] Fig. 1 (g) is a simplistic representation of the spin complex model. In the ground state, the two electrons of the spin pair are localised on Defect A in a closed shell spin singlet (S = 0) with an excited state triplet (S = 1) accessible by an intersystem crossing which leads to the $m_s = 0 \leftrightarrow \pm1$ transitions (green highlight).

Under optical excitation, an electron can undergo a charge transfer (red dashed arrow) onto Defect B, leaving both defects with one unpaired electron thus forming a weakly coupled spin pair which is well described by a mixed set of singlet and triplet states (S = ), and gives rise to the S = 1/2 transition (blue highlight). This is similar to the radical-pair mechanism well known in spin chemistry[28,29]. Since PL spectra of the quantum emitters consist of one strong ZPL, it is reasonable to assume only one of the defects is optically active or emits outside of our detection window in the visible range (550-1000 nm). Here we assume the emission occurs entirely from a localised transition at Defect A while it holds both electrons. Further detailed discussion on the potential level structure and possible defects in hBN is presented towards the end of the manuscript.

We now conduct a detailed analysis of the optically active quantum spin defects formed in oxygen annealed c-hBN flakes. PL from nine ODMR active emitters are shown in Fig. 2 (a), exhibiting narrowband emissions from 589 to 751 nm with distinct ZPLs at room temperature. The spread of ZPL emission is consistent with all emitters found in oxygen annealed c-hBN, offering no correlation between emission wavelength and ODMR activity. As such, magnetic field dependent PL or an ODMR measurement in the presence of the external magnetic field seems to be the only two methods to distinguish between ODMR active and inactive defects.

Fig. 2 (b-f) presents a selection of ODMR spectra. The majority of ODMR active emitters exhibit the characteristic three transitions, with the exception of (d), which displays only the S = 1/2 transition. The solitary S = 1/2 spin system is similar to previously studied emitters in hBN nanopowder and exfoliated flakes with no observable zero field splitting (ZFS)[7,17,19]. As mentioned above, while emitters with only the S = 1/2 transition were found, no emitters with only the S = 1 system were observed. Of the S = 1 transitions measured, the average ZFS parameters are D ≈ 960 ±110 MHz and E ≈ 70 ±40 MHz (See discussion on Fig. S8 for further details).

For comparison, we also plot in Fig. 2 (g) an example of an ODMR active emitter found in oxygen annealed pristine hBN. The ZPL characteristics are similar, as well as the corresponding ODMR spectrum that retains the characteristic S = 1 and S = 1/2 transitions seen in oxygen annealed c-hBN.

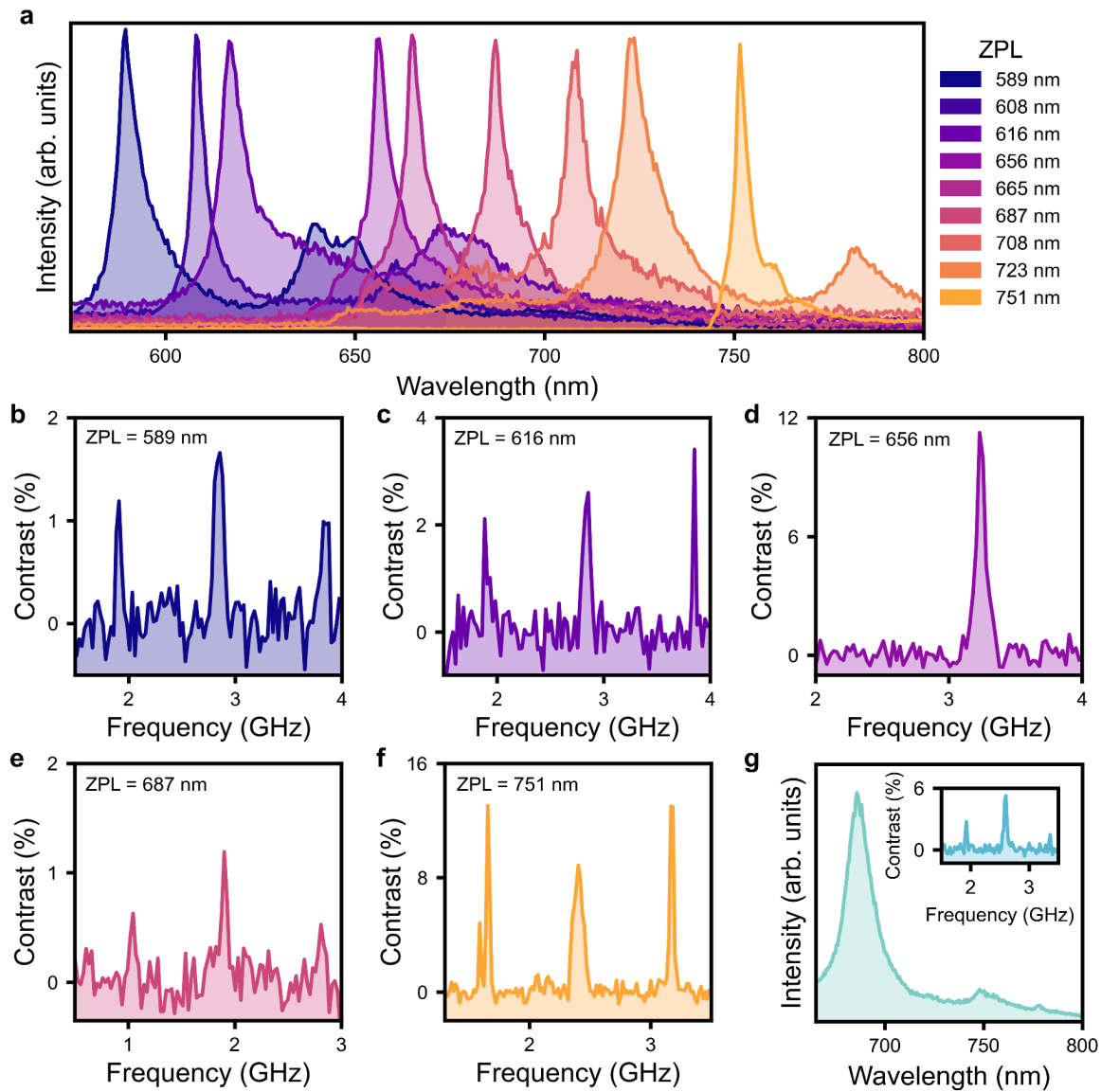

*Figure 2. Room temperature spectral characteristics of oxygen annealed emitters in c-hBN a) nine characteristic PL spectra of ODMR active emitters in c-hBN with ZPLs ranging from 589 to 751 nm. b-f) Corresponding ODMR spectra of five emitters chosen from the above spectra. The three transitions can be seen in all but (d), which only shows the S = 1/2 transition. The spectra were taken over a range of magnetic fields from 67 - 115 mT. g) comparison of a PL and ODMR spectra (inset) of an emitter found in pristine hBN that was annealed in oxygen also exhibiting the S = 1 and S = 1/2 spin transitions.*

Next, we performed statistical analysis to study the effect of carbon and oxygen elements on the generation of ODMR active emitters in hBN. While the presence of carbon within the defect is widely acknowledged by the scientific community, our current results elucidate the importance of oxygen in generating ODMR active defects during the thermal process. Over 11 flakes of each type of hBN were measured for data presented in Fig. 3. In both cases, emitters with ZPL emissions across the visible range are found, with the majority of the ZPLs being around 580 nm as shown in Fig. 3 (a). Previous studies on the fabrication of single-photon emitters in hBN suggested ZPL emissions around 600 nm are the result of carbon-related defects while the presence of oxygen increases the density of emitters with ZPL above 700 nm[30-35]. Our study shows oxygen annealing of c-hBN results in a higher percentage of emitters above 700 nm in C-hBN. Out of ~100 emitters found on hBN, 8% have ZPL above 700 nm while for ~130 emitters in c-hBN, 16% have ZPL between 700-850 nm.

On average the emitter densities were similar for both hBN and c-hBN flakes after oxygen annealing. This is shown on the left panel of Fig. 3 (b), where on average ~6.5 emitters were located within a 30x30 um area of the hBN or c-hBN flakes, respectively. Within an individual flake, the population of emitters can be as low as 3 or as high as ~18 in the same area for hBN or c-hBN flakes (see Fig. S4). Interestingly, while there are no significant differences in the distribution of emitters in hBN and c-hBN, oxygen annealing of c-hBN results in a significant increase in the density of ODMR active emitters, marking nearly an order of magnitude improvement over previously reported results.

It has been reported that only a small fraction — typically less than 5% (1 in 20) — of the defects in hBN are spin-active and most of the previous reports show only S = 1/2 transitions[6]. However, our results demonstrate a substantially higher percentage of ODMR active emitters in c-hBN, with approximately 25% (5 in 20) of the emitters exhibiting ODMR signatures, as shown in the right panel of Fig. 3 (b). This indicates that our technique in annealing c-hBN in oxygen atmosphere is a reliable method to dramatically increase the probability of activating optically active spin defects that are inherent but rarely present in hBN flakes or films. Moreover, the quality of the flake is largely maintained which in combination with narrowband emission from ODMR active defects, makes it attractive for employing them in nanophotonics and spin based quantum applications.

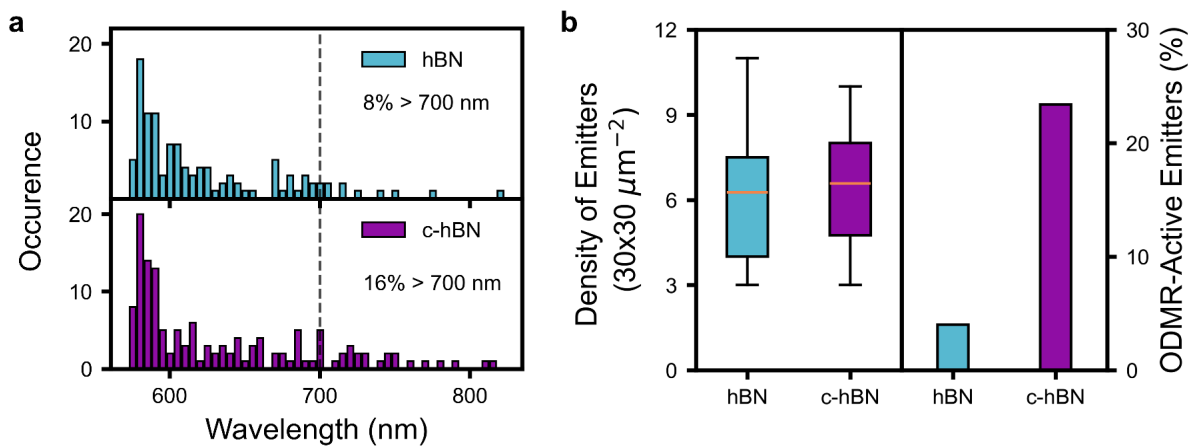

Figure 3. Statistical analysis of emitter wavelength, density and ODMR activity. a) Distribution of emitter ZPL wavelengths in 11 oxygen annealed hBN flakes (turquoise) and 11 c-hBN

*annealed flakes (purple), respectively. The vertical dotted line is a guide to the eye. b) Box plot of emitter density per 30x30 µm$^{-2}$ area for hBN and for c-hBN with mean value marked in orange (left panel). Percentage of emitters which are ODMR-active (right panel) for emitters in hBN and c-hBN flakes, respectively, after oxygen annealing.*

To verify the role of oxygen in the observed quantum emitters, we conducted a control experiment using argon as the annealing gas under identical experimental conditions (see Methods section). However, the results of the argon-annealed samples were significantly different from those of the oxygen-treated samples. Unlike the oxygen-annealed flakes, which displayed narrowband, well-defined emission from single-photon emitters, the argon-annealed samples predominantly exhibited broad, featureless emission spectra (Fig. S7). These broad emissions were often accompanied by a substantial PL background, which significantly impeded further characterization of any potential quantum emitters. Although oxygen annealing of hBN flakes produces a high density of single emitters, our result suggests the presence of both oxygen and carbon is necessary to increase the overall number of ODMR active emitters in hBN.

To build our understanding of the spin complex, we explore the electronic level structure by performing magnetic field dependence measurements. First, ODMR (at room temperature, from a single defect) was recorded at increasing out-of-plane magnetic field strengths, starting at 12 mT (Fig. 4 (a)). Optical characterisation of this emitter can be found in Fig. S10 (a-c). Dashed lines are calculated Zeeman shifts using the following equation along with the zero field values that best describe each transition.

$$\Delta E = \gamma_e B m_s \qquad (1)$$

Where ΔE is the change in energy, $\gamma_e$ is the gyromagnetic ratio of an electron (28.025 MHz/mT) and B is the external magnetic field applied. $m_s$ describes the magnetic quantum number i.e. $m_s$ = 0, ±1/2, ±1. The pink dashed line follows the S = 1/2 transition with no ZFS and a g-factor of 2 (i.e. $\Delta m_s$ = ±1). The S = 1 transitions are indicated by yellow dashed lines and also have a g-factor of 2. For this particular emitter, D was calculated to be 950 MHz and E = 200 MHz. A final transition, indicated by the purple dashed line, is also present within the measured emitter.

This transition has a slope corresponding to a g-factor of 4 (i.e. $\Delta m_s$ = ±2), which suggests a double quantum transition (DQT) and was thus ascribed to $m_s$ = -1 ↔ +1 of the S = 1 state. The extrapolated zero field splitting of the DQT was found to be 200 MHz matching the E parameter, further supporting its relationship to the S = 1 system present in the spin complex. Note that emitters with and without the DQT have been identified within oxygen annealed c-hBN however, no correlation has been established between its presence and other emitter characteristics.

A notable decay in contrast is measured as the magnetic field is decreased in all observed transitions. While this is consistent with previous observations for the S = 1/2 transition[6], for typical S = 1 spin systems it would be expected that there would be observable resonances at zero field. In Fig. 4 (b) three cross sections taken from Fig. 4 (a) at 18, 43 and 66 mT (marked with grey dotted lines) as well as an ODMR measurement of the same emitter at 0 mT are shown. The individual ODMR spectra at the non-zero fields highlight the diminishing contrast as the external field decreases with clearly no resonances observed in the absence of a magnetic field.

The presence of the S = 1 transitions and the decay in contrast can be explained by considering a photodynamic model. In Fig. 4 (c), we propose an extension upon the previous spin pair model to include the excited state spin triplet in the level structure which is inherent to the system and provides a plausible explanation to our results. The excited state of the optically active Defect A may undergo an intersystem crossing to this triplet state where spin driving can occur leading to the observed S = 1 resonances. We confirm the triplet is a metastable state through a series of pulsed measurements, (Fig. S11) which we refer to as the "Local MS". The "Remote MS" describes the weakly coupled spin pair configuration of the spin complex in which an electron from Defect A undergoes a charge transfer to Defect B. We define the spin pair levels as three triplet states $T_+$, $T_-$ and $T_0$ as well as a degenerate singlet state S. The double ended arrows in the level diagram are color matched to the electron transitions in Fig. 4 (a). The S = 1 transitions seen in Fig. 4 (a,b) correspond to the yellow double ended arrow in the Local MS state while the S = 1/2 transition is represented by the pink arrows in the Remote MS. The Remote MS has two transitions that are equal in magnitude at any magnetic field strength. At low magnetic field strengths, $T_+$ and $T_-$ approach the $T_0$ and S states, leading to spin mixing that diminishes the contrast of the spin-1/2 transition. The observed reduction in the contrast of the S = 1 transitions can be explained by the primary decay pathway from the Local MS to the ground state passing through the Remote MS. For example, at low magnetic fields (< ~30 mT), when the electron transfers to Defect B, the spin information is lost due to the mixing of the Remote MS sublevels. We note this mixing should lead to the contrast of the 0 ↔ +1 and 0 ↔ -1 transitions in the spin triplet to decay at the same rate however, there is an apparent asymmetry in our measurements where the 0 ↔ -1 transition (and spin-1/2) decays differently to the 0 ↔ +1 transition. We attribute this to low efficiency of the MW amplifier at lower frequencies.

It is key to note that the optical emission occurs only from the excited to the ground state, localised on Defect A, while all spin dependent transitions require the charge transfer into Defect B. If Defect B is too far (or too close) from the optically active Defect A, only the optical emission but no ODMR would be observed. A photodynamic model of the spin complex level structure was also simulated which produced optical spin transitions that closely match with the experimental results seen in Fig. 4 (a). This further validates our understanding of the spin complex. A detailed discussion on the model can be found in the Supporting Information and Fig. S13.

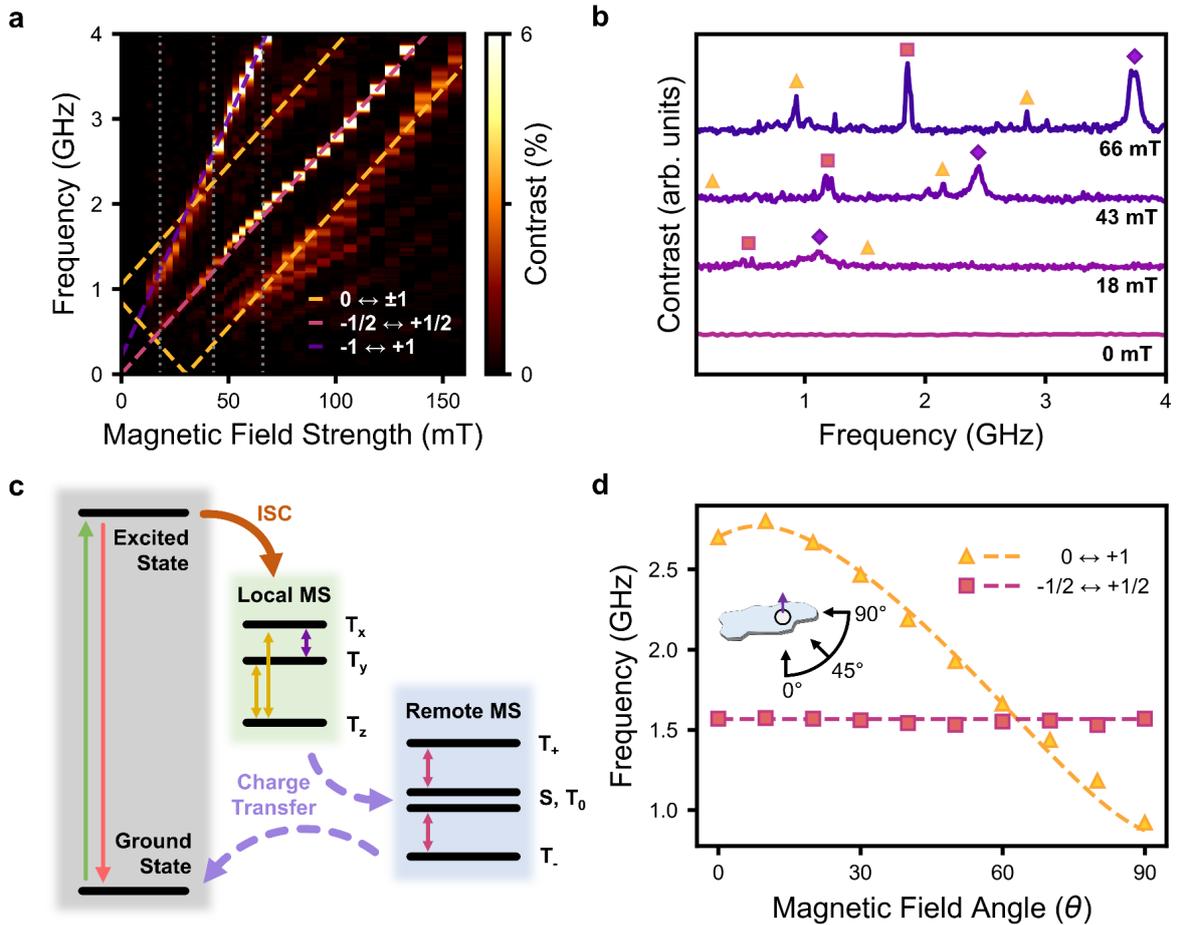

*Figure 4. Spin dynamics of the quantum emitters in hBN a) ODMR of a spin-active emitter under an out-of-plane magnetic field increasing from 12 to 160 mT. The colored dashed lines represent the calculated Zeeman shifts corresponding to potential spin transitions, the color legend indicates that yellow corresponds to the transition from 0 ↔ ±1, pink corresponds to the S = 1/2 transition, and purple corresponds to −1 ↔ +1. b) ODMR spectra taken from the three magnetic field strengths marked by grey dotted lines in a) as well as at 0 mT. Spin transitions are labeled with shapes of the same color as in (a). c) Proposed energy level diagram of the spin complex. Double ended arrows are color matched to the spin transitions of (a). The orange arrow labeled ISC depicts the intersystem crossing and dashed purple lines are the charge transfer pathways. $T_x$, $T_y$, $T_z$ and $T_+$, $T_-$, $T_0$ refer to levels within the two triplet states while S is a singlet within the Remote MS. d) Resonant frequency of the 0 ↔ +1 and -1/2 ↔ +1/2 transitions as a function of magnetic field angle, moving from out-of-plane (0°) to in-plane (90°) as illustrated by the inset. The dashed lines are simulated transition frequencies using Equation (1).*

Finally, to ascertain the spin quantization direction of the S = 1 system, ODMR was recorded for different field alignments. Further optical characterisation of this emitter can be seen in SI Fig. S10 (d-f). The shift for S = 1/2 and S = 1 transitions as a function of magnetic field angle (□) are shown in Fig. 4 (d) with pink squares and yellow triangles, respectively. Here, zero degrees corresponds to the magnetic field aligning with the out-of-plane axis (c axis) of the hBN flake as illustrated in the inset. As the magnetic field was rotated, the magnet moved away from the sample, causing the field strength magnitude to decrease slightly. This issue was corrected with a multiplying factor. With this correction, we find the experimental results to be in excellent agreement with simulations of a S = 1 spin Hamiltonian and determine an out-of-plane angle (±5°) for the 0 ↔ ±1 (S = 1) transitions. This is in contrast with the S = 1 single photon emitters reported in metal-organic vapour phase epitaxy hBN, where the in-plane nature and ZFS parameter D = 1.96 GHz, indicating a fundamentally different defect structure. The full data set and model can be seen in Fig. S12.

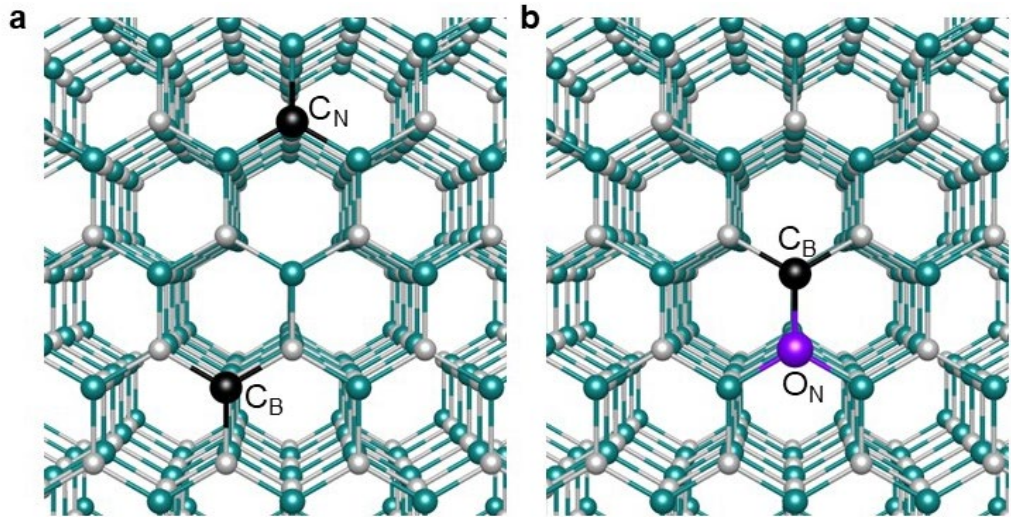

Figure 5. Potential defects for spin active emitters in hBN. a) Structure of $C_BC_N$ donor-acceptor pair (DAP-sqrt(13)) emitting at ~750 nm. b) Structure of the $C_BO_N$ defect possessing spin-1/2 ground state in its positive charge state. When both defects are in a close proximity, they form the spin complex.

Finally, we discuss the potential microscopic structures and the role of oxygen treatment in the formation of spin complexes in hBN. Our samples are assumed to already contain carbon-based emitters that emit across a broad spectral range. An appealing explanation for the diverse emission energies of carbon-based centers is the model of closely spaced donor-acceptor (DAP) complexes. For simplicity, we consider aggregates of $C_B$ and $C_N$ defects, denoted as $C_BC_N$-DAP-$X$ defects, where $X$ is a multiple of the B-N bond length. Among these, the calculated emission wavelengths of neutral $C_BC_N$-DAP-$\sqrt{13}$ (746 nm) and $C_BC_N$-DAP-$\sqrt{4}$ (784 nm) fit our experimental data most closely (Fig. 5 (a)). We note that other carbon based complexes (e.g. $C_BC_{2N}$ and $C_NC_{2B}$) were proposed[36-38], and are potentially viable to be considered as "Defect A" (the optically active defect). However, as experimental results suggest, oxygen plays an important role in activating spin based emitters in hBN.

To investigate the role of oxygen treatment in creating emitters and spin complexes, we perform density functional theory calculations on single-site $O_N$ and $O_B$ defects, as well as all possible combinations of $O_N$, $O_B$, $C_N$, and $C_B$ defects in various charge states. Interestingly, we do not find intra-defect state transitions that could directly enhance emission in longer than 700 nm wavelength range. Instead, we conclude that the observed increase in emitters is likely due to the activation of the bright state of carbon-related centers in hBN. Oxygen treatment can raise the Fermi energy toward the middle of the band gap, through the formation of $O_N$ donor defects [22]. Neutral $C_BC_N$ DAP pairs with larger separations, such as DAP-sqrt(13) and DAP-4, are stable only within a narrow Fermi energy range near the middle of the band gap. Thus, the incorporation of $O_N$ defects and pinning the Fermi-level at the middle of the band gap stabilises the neutral charge states of these pairs, leading to emissions at >700 nm.

Furthermore, regarding the dark spin counterparts of spin complexes, we conclude that $O_N$ defects cannot directly contribute to ODMR signals. In their spin-1/2 neutral charge state, these defects exhibit a strong hyperfine interaction with a boron atom neighboring the oxygen site. The most likely oxygen-related spin defect contributing to ODMR-active spin complexes is the $C_BO_N$ complex which can act as a donor and possesses a spin-1/2 ground state in the positive charge state ($C_BO_N^+$) (Fig. 5 (b)). Our model for the observed signals involves a spin-0 $C_BC_N$-DAP-√(13) (or DAP-4) color center paired with a spin-0 $C_B$-$O_N$ complex. In the metastable state, the DAP becomes charged (spin-1/2), while the $C_B$-$O_N$ defect transitions to its positive (spin-1/2) charge state. These two spin-1/2 states produce the doublet resonance, while spin-flip transitions in the $C_BC_N$ DAP lead to local metastable state with triplet resonances. Thus, this combination offers a solid explanation to why we always only observe S=1 signal associated with a S=1/2 system in all hBN flakes.

In summary, our work provides a solid groundwork for activation of narrowband single photon emitters in hBN across the visible range which are spin active. These spin complex emitters have both S = 1 and S = 1/2 transitions and their density is increased by more than order of magnitude following annealing c-hBN in the oxygen atmosphere. Moreover, for the first time we show a clear protocol to achieve high density of spin defects in exfoliated flakes (rather than polycrystalline films or powders) that is significant for future photonic integration. The optical spin readout mechanism of these emitters are explained by considering two defects including an optically active DAP pair which agrees with having ZPLs across all the visible range. Through DFT calculation, a potential defect structure is proposed based on optical emission from $C_BC_N$-DAP-*X* with nearby dark $C_BO_N^+$. All in all, our work serves as a cornerstone for further development of integrated quantum photonic devices employing hBN and other van der Waals crystals and facilitates further interest into fundamental research on spin complex systems.

Note: During the preparation of our manuscript, we became aware of a similar work [39], that reported generation of single spin defects by carbon implantation into exfoliated hBN flakes, and similar interplay of S=1 and S=1/2 defects. Our results are complementary in terms of methodology to achieve a similar outcome.

**METHODS**

Sample Preparation: hBN crystals and carbon-doped hBN crystals are mechanically exfoliated using scotch tape, flakes are then transferred onto a $SiO_2$ substrate using Gel-Pak on 80°C hotplate. The samples are placed in a quartz boat and annealed in a quartz tube on the Lindberg Blue 3000 Furnace with an oxygen flow rate of 1000 sccm under vacuum ~ 30 mTorr for 4 hours at 1000°C. During cool down the gas flow is changed to argon at 400°C. The samples are then placed into a UV-Ozone for 4 hours before measuring.

Optical Measurements: Photoluminescnece measurements were performed using a 532 nm continuous wave laser and a high 0.9 NA 100 x objective (Olympus) with a 568 nm long-pass filter in place for the laser and the samples are mounted on a piezo stage. For correlation measurements, the collected light is directed into a Hanbury Brown and Twiss (HBT) interferometer with two avalanche photodiodes (APD) (Excelitas), the data acquisition are measured by a coincidence counter module (PicoQuant).

Optically Detected Magnetic Resonance (ODMR): To perform ODMR, a few extra pieces of hardware were added to the optical setup. For most measurements, a sputtered copper antenna provided the radio frequency (RF) signal for driving the spin transitions. This antenna was fabricated using UV photolithography, with a 5 nm chromium adhesion layer, followed by 500 nm of copper to form the antenna. A few hBN flakes could then be aligned-transferred on top of the antenna. Due to the high-temperature annealing required for emitter creation, the flakes become strongly bonded to the substrate and polyvinyl alcohol (PVA)-assisted transfers proved insufficient for pickup. Consequently, a polyethylene terephthalate (PET) stamp which provides a higher level of adhesion was employed with positive results (See aligned-transfer section below). Some measurements were achieved by simply suspending a copper wire (50 μm diameter) over the sample, removing the need for flake transfers.

For constant wave (CW) ODMR, an RF on/off sweep of frequencies was implemented. A signal generator (AnaPico APSIN 4010) steps through the range of frequencies, which are then amplified (Minicircuits ZHL-16W-43-S+), with each step consisting of 1 ms of RF signal followed by a 1 ms wait time to allow for a reference measurement. For each step the ODMR contrast percentage is calculated using the equation: Contrast (%) = (Signal-Reference)/Reference * 100. An acousto-optical modulator in the laser path as well as an RF switch (Minicircuits ZYSWA-2-50DR+) are controlled by a SpinCore PulseBlaster to achieve the pulsed ODMR measurements. The external magnetic field is achieved with a NdFeB magnet (N35) that is positioned below the sample stage and can be raised and lowered with a Z-axis stage. The magnet can also be placed on a rotating mount to provide variation in magnetic field angle.

Polyethylene Terephthalate (PET) Aligned-Transfer: A 2mm x 2mm piece of polyethylene terephthalate (PET) was attached to a microscope slide using double sided tape to act as the aligned-transfer stamp. The PET is then checked for bubbles or contaminants, and if none are present, both the starting sample substrate and stamp are positioned on a home-built aligned-transfer setup with two XYZ stages. Once the desired flake is found on the substrate, it is heated to 70°C and the PET stamp is brought down onto the flake. After one minute, the flake can be lifted up. To place the flake down, swap the starting substrate for the target substrate and bring the PET down on it, aligning the flake with the area of the substrate you wish to place the flake. Heat to 80°C then heat to a further 130°C at 15°C/min. At this temperature, the double sided tape breaks down and the PET stamp can be removed from the microscope slide. Cool down to 70°C and leave the sample in dichloromethane (DCM) for up to 12 hours, or until PET is dissolved.

**Acknowledgements**

We acknowledge financial support from the Australian Research Council (CE200100010, FT220100053, DP240103127, FT200100073 and DP250100973), and the UTS node of the ANFF for access to nanofabrication facilities. I.O.R is supported by an Australian Government Research Training Program Scholarship.

**Author contributions**
MK and IA designed the experiments. BW, HZJZ and MK. wrote the manuscript with contributions from all co-authors. HZJZ, BW and JL-W performed experimental measurements and data analysis. IR, J-PT and VI performed the computational calculations. TT and KW performed hBN and c-hBN growth. BW and MK performed ODMR experiments. MT, IA and MK supervised the project. Co-first authorship order was determined via a single coin toss. Both BW and HZJZ contributed equally. All authors discussed the results and contributed to the manuscript.

**Competing interests**
The authors declare no competing interests.